\documentclass[aps,prd,superscriptaddress,floatfix,nofootinbib,preprintnumbers]{revtex4}
\usepackage{amsmath,multirow}
\usepackage{graphicx,tabularx}
\usepackage{url}
\usepackage{footmisc}

\newcommand{\sps}{SPS1a${}_{1000}$}

\newcommand{\sul}{\ensuremath{\tilde{u}_L}}
\newcommand{\sdl}{\ensuremath{\tilde{d}_L}}
\newcommand{\ssl}{\ensuremath{\tilde{s}_L}}
\newcommand{\scl}{\ensuremath{\tilde{c}_L}}
\newcommand{\sur}{\ensuremath{\tilde{u}_R}}
\newcommand{\sdr}{\ensuremath{\tilde{d}_R}}
\newcommand{\ssr}{\ensuremath{\tilde{s}_R}}
\newcommand{\scr}{\ensuremath{\tilde{c}_R}}

\newcommand{\sq}{\ensuremath{\tilde{q}}}
\newcommand{\su}{\ensuremath{\tilde{u}}}
\newcommand{\sd}{\ensuremath{\tilde{d}}}

\newcommand\one{\leavevmode\hbox{\small1\normalsize\kern-.33em1}}

\newcommand{\qqquad}{\qquad \qquad}

\newcommand{\matx}{|\mathcal{M}|^2}

\newcommand{\msbar}{\ensuremath{\overline{\text{MS}}}}
\newcommand{\drbar}{\ensuremath{\overline{\text{DR}}}}

\newcommand{\lqcd}{\Lambda_\text{QCD}}

\newcommand{\go}{\tilde{g}}

\newcommand{\nz}[1]{\tilde{\chi}_{#1}^0}
\newcommand{\cpm}[1]{\tilde{\chi}_{#1}^\pm}

\providecommand{\mgo}{m_{\tilde{g}}}

\providecommand{\msq}{m_{\tilde{q}}}

\newcommand{\mne}[1]{m_{\tilde{\chi}^0_{#1}}}

\newcommand{\gev}{{\ensuremath\rm GeV}}
\newcommand{\tev}{{\ensuremath\rm TeV}}
\newcommand{\fb}{{\ensuremath\rm fb}}

\def\slashchar#1{\setbox0=\hbox{$#1$}           % set a box for #1
   \dimen0=\wd0                                 % and get its size
   \setbox1=\hbox{/} \dimen1=\wd1               % get size of /
   \ifdim\dimen0>\dimen1                        % #1 is bigger
      \rlap{\hbox to \dimen0{\hfil/\hfil}}      % so center / in box
      #1                                        % and print #1
   \else                                        % / is bigger
      \rlap{\hbox to \dimen1{\hfil$#1$\hfil}}   % so center #1
      /                                         % and print /
   \fi}

\def\eg{{\sl e.g.} \,}
\def\ie{{\sl i.e.} \,}

\begin{document}

\date{\today}

\title{Automized Squark--Neutralino Production to Next-to-Leading Order}

\author{Thomas Binoth\footnote{deceased}}
\affiliation{SUPA, School of Physics \& Astronomy, The University of Edinburgh, UK}

\author{Dorival Gon\c{c}alves Netto}
\affiliation{Institut f\"ur Theoretische Physik, Universit\"at Heidelberg, Germany}

\author{David L\'opez-Val}
\affiliation{Institut f\"ur Theoretische Physik, Universit\"at Heidelberg, Germany}

\author{Kentarou Mawatari}
\affiliation{Theoretische Natuurkunde and IIHE/ELEM, Vrije Universiteit Brussel, Belgium}
\affiliation{International Solvay Institutes, Brussels, Belgium}

\author{Tilman Plehn}
\affiliation{Institut f\"ur Theoretische Physik, Universit\"at Heidelberg, Germany}

\author{Ioan Wigmore}
\affiliation{SUPA, School of Physics \& Astronomy, The University of Edinburgh, UK}

\begin{abstract}
  The production of one hard jet in association with missing
  transverse energy is a major LHC search channel motivated by many
  scenarios for physics beyond the Standard Model. In scenarios with a
  weakly interacting dark matter candidate, like supersymmetry, it
  arises from the associated production of a quark partner with the
  dark matter agent. We present the next-to-leading order cross
  section calculation as the first application of the fully automized
  {\sc MadGolem} package. We find moderate corrections to the production
  rate with a strongly reduced theory uncertainty.
\end{abstract}

\maketitle

%%%%%%%%%%%%%%%%%%%%%%%%%%%%%%%%%%%%%%%%%%%%%%%%%%%%%%%%%%%%%%%%%%%%%%%%
\section{Introduction}
\label{sec:intro}

Since the LHC started running at a center-of-mass energy of 7~TeV
searches for new physics are a major effort, realized in a rapidly
increasing number of publications~\cite{review}.  Inclusive searches
for supersymmetry at the LHC have started to constrain the allowed
parameter space of the minimal supersymmetric Standard
Model~\cite{susy_lhc}, most notably in the part of the squark--gluino
mass plane which can be described in terms of gravity mediation. Such
searches are based on jet production from squark and gluino decays and
two stable lightest supersymmetric particles (LSP). The latter could be
a dark matter agent with a weak-scale mass.

\begin{figure}[b]
\includegraphics[width=0.4\textwidth]{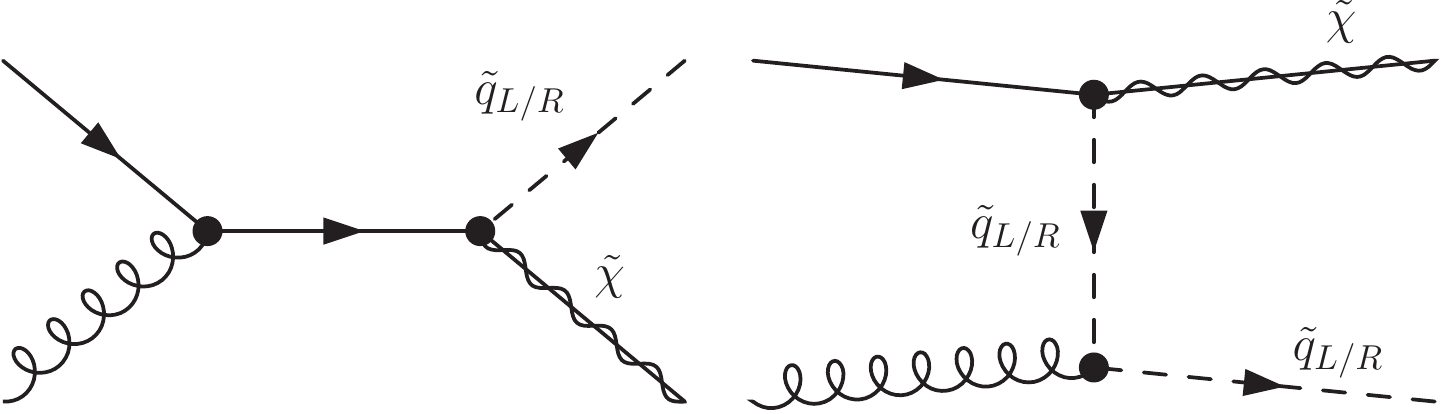} 
\caption{Feynman diagrams for the associated production of a squark
  and a gaugino to leading order.}
\label{fig:feyn1}
\end{figure}

The main production mode for jets and dark matter particles at the LHC
would most likely be squark or gluino pair production, mediated by the
strong interaction~\cite{autofocus}. The limitation of this channel is
that it will be hard to extract any model parameters beyond the masses of
the new particles~\cite{sfitter_fittino}. The production is
governed by the strong interaction and the (sum of) branching ratio(s)
leading to jets plus missing transverse energy can be expected to be
close to unity. Therefore, it is worth studying additional production
processes which directly involve the weakly interacting sector of the
new physics model. In supersymmetry, those are the associated
production of a gluino~\cite{gluino_asso} or a squark with a
neutralino or chargino~\cite{squark_asso}
\begin{alignat}{5}
 pp \; \to \; \sq \nz, \quad \sq \tilde{\chi}^\pm \; .
\end{alignat}
The leading order Feynman diagrams for this process we show in
Fig.~\ref{fig:feyn1}. This channel naturally leads to one single hard
decay jet and missing energy. This signature is not unique to
supersymmetry or other models with quark partners and a weakly
interacting dark matter agent; it also constitutes the theoretically
most reliable signature for large extra dimensions~\cite{extrad}. In
that sense, observing single jet production with missing energy would
be one of the most exciting anomalies to interpret at the LHC.\bigskip

%------------------------------------------------
%------------------------------------------------

Aside from the quark-gluon and squark-gluon QCD vertices, the leading-order process is driven
by the $q$-$\sq$-$\tilde{\chi}$ interaction. Because the dominant
light-flavor quarks only have a tiny Yukawa coupling, this interaction
relies on the two weak gauge charges of the quark-squark pair
involved. This way, it carries information on the composition of the
dark matter candidate $\nz{1}$ and an accurate measurement would also
allow improved predictions for the direct detection and relic density
of dark matter.  Finally, this process probes the supersymmetric
relations between the gauge couplings and the gaugino-quark-squark
couplings. Prospects for carrying out such a measurement out of the
mono-jet rates at the LHC have been studied in
Ref.~\cite{Allanach:2010pp}.\bigskip
  
Next-to-leading order QCD contributions to this production process
arise at order $\alpha \alpha_s^2$ and originate from both QCD (gluon
mediated) and SUSY-QCD (gluino mediated) contributions. They involve
virtual one-loop corrections as well as real emission off the initial
state partons or the final state squark. While we expect the size of
the virtual corrections combined with real emission to stay moderate,
there are potential sources of huge corrections; for example, the
production of a squark pair where one of the two squarks decays to a
quark and a neutralino contributes at the same order $\alpha
\alpha_s^2$, but can effectively be an overwhelming two particle
production process. To avoid double counting and instead allow for a
clear separation and prediction of the two channels, the treatment of
on-shell singularities in associated squark and neutralino production
is crucial. Note that this separation without double counting serves a
theoretical purpose. For an actual observable, we always need to
combine QCD pair production with associated production because initial
state jet radiation at the LHC can be as hard as decay
jets~\cite{qcd_radiation} and cannot be distinguished event by
event.\bigskip

Finally, we use this production process to show how heavy
particle pairs at next-to-leading order can be described by the fully
automized tool {\sc MadGolem}. It generates all tree-level diagrams
and the corresponding helicity amplitudes in the {\sc Madgraph}
framework~\cite{madgraph}, based on {\sc Helas}~\cite{helas}. The
virtual corrections are generated with {\sc Qgraf}~\cite{qgraf} and
numerically computed by {\sc Golem}~\cite{golem}. Supersymmetric
counter terms are part of the model implementation. On-shell
subtraction terms in the {\sc Prospino}
scheme~\cite{prospino_squark,thesis,prospino_onshell} can also be
generated automatically. Further details we provide in the appendix. 
To our knowledge, this process is the first
fully automized beyond-the-Standard-Model next-to-leading order (NLO) computation based on a
(soon-to-be) public add-on to a major Monte Carlo generator, meant to
be used independently by the LHC experimental community.

%%%%%%%%%%%%%%%%%%%%%%%%%%%%%%%%%%%%%%%%%%%%%%%%%%%%%%%%%%%%%%%%%%%%%%%%
\section{S-up production in association with the LSP}
\label{sec:nlo}

In this first section we focus on total rates for $\sur
\nz{1}$ and $\sul \nz{1}$ production and features characterizing the
NLO effects. The results we present in terms of the consistent ratio
$K = \sigma^\text{NLO}/\sigma^\text{LO}$, an approach which we will
modify once we study distributions in Sec.~\ref{sec:distri}.

To avoid the current LHC bounds on squark and gluino
production~\cite{susy_lhc} we use a modified SPS1a~\cite{sps}
point (\sps) where the gluino mass is increased to
$1~\tev$. Because the gluino does not appear in the LO
Feynman diagrams, its increased mass will merely reduce the impact of SUSY-QCD
corrections which are nevertheless fully included --- some effective 
theory issues with the proper decoupling of the gluino we 
discuss in detail later.\bigskip

In our numerical analysis we use the CTEQ6L1 and CTEQ6M parton
densities with five flavors~\cite{cteq}.  For the strong coupling we
consistently rely on the corresponding $\alpha_s(\mu_R)$. Its value we
compute using two-loop running from $\lqcd$ to the required
renormalization scale, again with five active flavors. For the central
renormalization and factorization scales we use the average final state mass $\mu_R^0
= \mu_F^0 = (\msq + m_{\tilde{\chi}})/2$, which has been shown to
lead to stable perturbative
results~\cite{prospino_squark,prospino_chargino}.

%------------------------------------------------
\begin{table}[b]
\begin{small}
\begin{tabular}{r||c|r|r|c||r|r|r|c||c|c} \hline
$\sqrt{S} \, [\tev]$ && $\sigma^\text{LO}$ [\fb] & $\sigma^\text{NLO}$ [\fb] & $K$
&& $\sigma^\text{LO}$ [\fb] & $\sigma^\text{NLO}$ [\fb] & $K$  & $m_{\tilde{q}_R} \,[\gev]$ & $m_{\tilde{q}_L} \,[\gev]$   \\ \hline 
7 & \multirow{2}{*}{$\sur\,\nz{1}$}& 29.62 & 42.17& 1.42 
&\multirow{2}{*}{$\sul\,\nz{1}$} &0.83 & 1.26 & 1.52 &  \multirow{2}{*}{549} &  \multirow{2}{*}{561} \\ 
14  && 176.36 & 245.74 & 1.39 & 
& 5.03 & 7.52 & 1.49 & \\ \hline
7 & \multirow{2}{*}{$\sdr\,\nz{1}$} &3.61 & 5.31 & 1.47 
&\multirow{2}{*}{$\sdl\,\nz{1}$} &1.21 & 1.77 & 1.46 &  \multirow{2}{*}{545} &  \multirow{2}{*}{568}  \\  
14 && 24.89 & 35.50 & 1.43 & 
& 8.67 & 12.37 & 1.43 & \\ \hline
7 & \multirow{2}{*}{$\scr\,\nz{1}$} &1.12 & 1.81& 1.61 
&\multirow{2}{*}{$\scl\,\nz{1}$} &0.03& 0.06 & 2.00 &  \multirow{2}{*}{549}  &  \multirow{2}{*}{561} \\ 
14 && 13.69 & 20.69 & 1.51 &
& 0.38 & 0.66 & 1.70 & \\ \hline
7 & \multirow{2}{*}{$\ssr\,\nz{1}$} &0.57& 0.78 & 1.38 
&\multirow{2}{*}{$\ssl\,\nz{1}$} &0.19 & 0.29 & 1.56 &  \multirow{2}{*}{545}  &  \multirow{2}{*}{568} \\
14 && 5.86 & 8.45 & 1.44 & 
& 2.00 & 2.98 & 1.49 & \\ \hline
7 & \multirow{2}{*}{$\sum \sq_{R}\,\nz{1}$} & 34.92& 50.07 & 1.43 &
\multirow{2}{*}{$\sum \sq_{L}\,\nz{1}$} & 2.26 & 3.38 & 1.50 \\
14 && 220.80 & 310.38 & 1.41 && 16.08 & 23.53 & 1.46 \\ \hline
\end{tabular}
\end{small}
\caption{Individual production rates $\sigma(pp \to \sq \nz{1})$ and corresponding $K$ factors for
  the modified \sps\, scenario. The first and second generation squark
  masses happen to be degenerate. The scales are set to their central
  values $\mu_R^0 = \mu_F^0 = (\msq + \mne{1})/2$. In the last
  line we show the sum of all contributions.}
\label{tab:channels}
\end{table}
%------------------------------------------------

Before we discuss the features of the NLO corrections
in detail for the (dominant) $\su \nz{1}$ production channel, in
Tab.~\ref{tab:channels} we quote the individual cross sections for all
light-flavor $\sq \nz{1}$ channels.  

The results in Tab.~\ref{tab:channels} clearly reflect the
flavor-locked nature of the process; ordered by the squark flavor in
the final state, all contributions stemming from sea quarks are
essentially irrelevant.  Our single jet signature is driven by the
$\tilde{u}$ and $\tilde{d}$ contributions, with the second generation
contributing at the $5\%$ level and, as we will see, within the NLO
scale uncertainty. Bottom-induced sbottom production we expect to be
further suppressed even though moderately large collinear logarithms
might enhance such a signature for very light sbottoms.

In addition, the modified SPS1a parameter point is a fairly generic
scenario for the weak sector and accommodates a relatively light
mostly bino LSP.  As a consequence, the neutralino coupling strength
to the right and left squarks is substantially different, \ie
$g_{u\sul\nz{1}}/g_{u\sur\nz{1}} \simeq 0.176 \sim 1/6$.  This explains the
difference between the LO cross sections for $\sul\nz{1}$ and
$\sur\nz{1}$ production of roughly one order of magnitude. As a bottom
line, we see that for a bino LSP more than 80\% of the total
squark-LSP production rates comes from the $\sur\nz{1}$
contribution.

%%%%%%%%%%%%%%%%%%%%%%%%%%%%%%%%%%%%%%%%%%%%%%%%%%%%%%%%%%%%%%%%%%%%%%%%
\subsection{Real and virtual corrections}

%------------------------------------------------
\begin{figure}[b]
\includegraphics[width=0.60\textwidth]{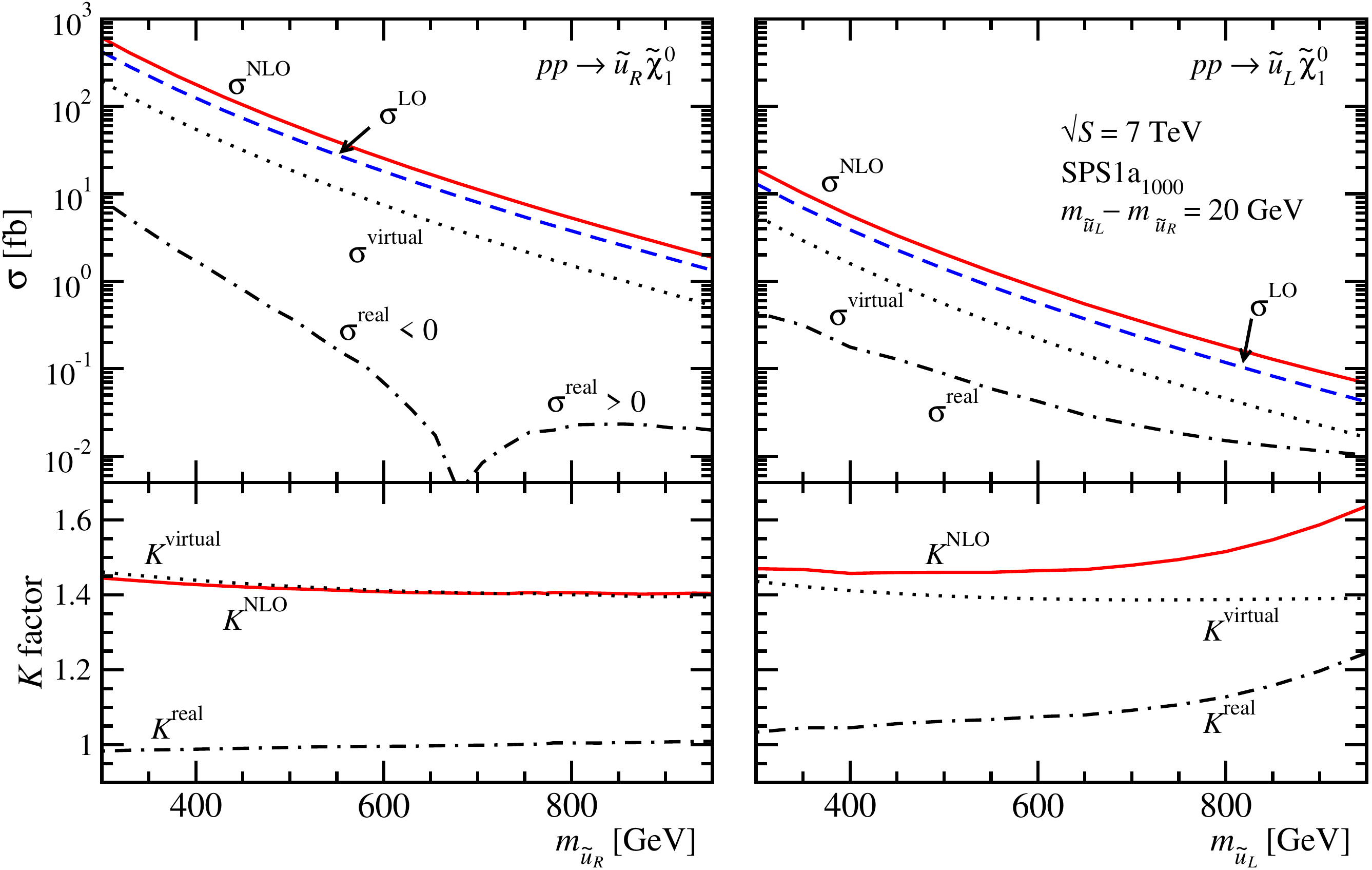} 
\caption{Cross sections $\sigma(pp \to \tilde{u}_{R/L} \nz{1})$ (top
  panels) and $K$ factor (bottom panels) as a function of $m_{\tilde{u}_{R/L}}$
  assuming $m_{\sul} - m_{\sur} = 20\,\gev$. For negative
  contributions to the total rate we show $|\sigma|$. The remaining
  MSSM parameters are fixed to our benchmark point. Real and the
  virtual corrections are separated using the original Catani-Seymour
  dipoles~\cite{catani_seymour} with $\alpha=1$, the integrated
  dipoles are included in the virtual corrections. The LHC energy is
  7~TeV.}
\label{fig:overmass}
\end{figure}
%------------------------------------------------

In Fig.~\ref{fig:overmass} we show the total cross sections as a
function of the squark mass. To assess the relative impact of the real
emission versus virtual corrections, and to spell out the differences
between the right and left squarks, we only show results for
$\sur\nz{1}$ and $\sul\nz{1}$ production. The former drives the bulk
of the overall $\sq\nz{1}$ event rate.  The two up squark masses we
vary simultaneously with a fixed splitting $m_{\sul} - m_{\sur} =
20\,\gev$.  As we can see, the virtual corrections are the dominant
NLO effects, leading to a NLO correction of the order $K\sim
1.4$.\bigskip

As real corrections we consider all contributions to the NLO cross
section with a three-particle final state. Virtual corrections include
gluon and gluino induced loops, but also integrated dipoles following
the Catani-Seymour dipole prescription~\cite{catani_seymour}. While
the dipole subtraction always covers the soft and collinearly
divergent phase space regions, in terms of a variable parameter
$\alpha$ they can be defined to extend more ($\alpha=1$) or less
($\alpha \ll 1$) into the non-divergent phase space
regime~\cite{alpha}. Unlike for the distributions shown in
Sec.~\ref{sec:distri} in this section we use $\alpha=1$, as in the
original implementation.

\begin{figure}[t]
\includegraphics[width=0.7\textwidth]{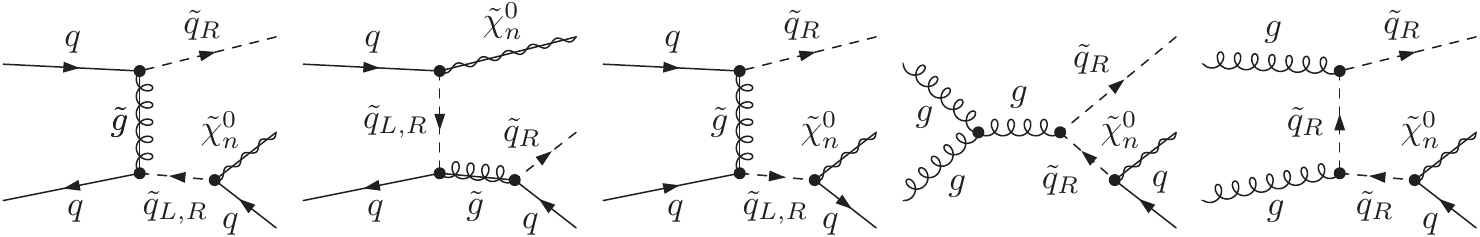} 
\caption{Example Feynman diagrams for real emission corrections to
  $\sq_R \nz{}$ production via intermediate on-shell states.}
\label{fig:feyn2}
\end{figure}

Given this choice, in Fig.~\ref{fig:overmass} we see that
real corrections are generally small compared to their virtual
counterparts. They exhibit an interesting feature, namely a positive
(negative) contribution for the $\sul\nz{1}$ ($\sur\nz{1}$) channel.
We can understand this feature through the different couplings to the
neutralino: real corrections to the $\sul\nz{1}$ channel mainly depend
on $g_{u\sul\nz{1}}$, but at next-to-leading order also
$g_{u\sur\nz{1}}$ contributes. For example, this happens for real
corrections triggered by the $gg, uu$ and $u\bar{u}$ initial states,
as shown in Fig.~\ref{fig:feyn2}.  This way, the NLO production rates
no longer factorize with the LO quark-squark-neutralino coupling.
Such higher order effects are sensitive to the coupling imbalance at
our parameter point, recalling $g_{u\sur\nz{1}}/g_{u\sul\nz{1}} \simeq
6$ owing to the bino-like nature of the LSP dark matter
candidate. This is why the real corrections to the $\sur\nz{1}$
channel are much smaller, and of the opposite sign than for the $\sul
\nz{1}$ channel.

%------------------------------------------------
%------------------------------------------------

%%%%%%%%%%%%%%%%%%%%%%%%%%%%%%%%%%%%%%%%%%%%%%%%%%%%%%%%%%%%%%%%%%%%%%%%
\subsection{Renormalization}
\label{sec:renorm}

Before we describe the {\sc MadGolem} combination of the virtual and
real corrections we need to ensure that ultraviolet and infrared
divergences are properly separated.  A crucial ingredient to our
higher order calculation therefore is the automatic treatment of the
ultraviolet divergences, \ie renormalization. In the partonic LO
cross section there appear three parameters which require
renormalization in QCD or SUSY-QCD. First, the squark mass is
renormalized in the on-shell scheme, which means that all
contributions from gluon or gluino exchange are absorbed into the mass
counter term. For QCD corrections this scheme can be fully extended to
mixing squarks~\cite{stop_dec,thesis}, but in this work we assume the
squark mixing angle to lead to negligible effects.  Second, we
renormalize the strong coupling constant in the five-flavor $\msbar$
scheme and explicitly decouple the heavier colored particles from its
running. This zero-momentum subtraction
scheme~\cite{prospino_squark,Berge:2007dz} leaves the renormalization
group running of $\alpha_s$ to be determined only by the lightest
colored particles. It corresponds to the definition of the measured
value of the strong coupling, for example in a combined fit with the
parton densities. The renormalization constant then reads
\begin{alignat}{5}
\delta\,Z_{g_s} = 
- \frac{\alpha_s}{4\pi} \;
  \frac{\beta^L_0+\beta^H_0}{2} \, \Delta_\epsilon
%+ \frac{\beta^H_0}{2} \, \Delta_{\epsilon}^\text{IR}
- \frac{\alpha_s}{4\pi} \,
\left( \frac{1}{3} \, \log \frac{m^2_t}{\mu_R^2}
      + \log \frac{m^2_{\go}}{\mu_R^2}
      + \frac{1}{12} \, \sum_\text{12 squarks} \,
        \log \frac{m^2_{\sq_j}}{\mu_R^2}
\right) \; ,
\label{eq:alphas_ct}
\end{alignat}
in terms of $\Delta_\epsilon = 1/\epsilon - \gamma_E + \log(4\pi)$.
%The labels correspond to an infrared or ultraviolet divergence,
%in the sense of the massless scalar two point function being
%$B(0,0,0) \propto \Delta_\epsilon^\text{UV}-\Delta_\epsilon^\text{IR}$.
The coefficient of the QCD beta function we can decompose
into contributions from light and heavy particles, 
$\beta_0 = \beta^{L}_0 + \beta_0^{H}$, with
\begin{equation}
\beta_0^L = \left[\frac{11}{3}\,C_A - \frac{2}{3}\,n_f \right]
\qqquad \text{and} \qquad 
\beta_0^H = \left[- \frac{2}{3}  -\frac{2}{3}\,C_A - \frac{1}{3}\,(n_f+1) \right] \; . 
\end{equation}
The number of active flavors is $n_f = 5$.
Correspondingly, the gluon field renormalization constant can be
written as
\begin{alignat}{5}
\delta\,Z_G = 
- \frac{\alpha_s}{4\pi} \;
  \left(\beta^L_0 + \beta^H_0 \right) \, 
%\left(\Delta_\epsilon^\text{UV}-\Delta_\epsilon^\text{IR} \right)
\Delta_\epsilon
+  \frac{\alpha_s}{2\pi}\,
\left( \frac{1}{3} \, \log \frac{m^2_t}{\mu_R^2}
      +\log \frac{m^2_{\go}}{\mu_R^2}
      +\frac{1}{12} \, \sum_\text{12 squarks} \,
       \log \frac{m^2_{\sq_j}}{\mu_R^2}
\right)
\label{eq:gluon_ct}.
\end{alignat}
Reflecting the underlying Slavnov-Taylor identities, the finite parts
of both renormalization constants are related as $\delta\,Z_G =
-2\,\delta\,Z_{g_s}$.

Finally, we need to compensate for dimensional regularization and the
$\msbar$ scheme breaking supersymmetry through the mismatch of two
gaugino and $D-2 = 2-2\epsilon$ gauge vector degrees of
freedom~\cite{Martin:1993yx,signer_stoeckinger}.  This is done by a
finite SUSY-restoring counter term in the quark-squark-gaugino Yukawa
coupling $g_{q\sq\tilde{\chi}}$ with respect to the associated gauge
coupling. For the $q$-$\sq$-$\tilde{\chi}$ coupling, this prescription
translates into a shift in $g_2 = e/s_w$~\cite{prospino_chargino}. The
SUSY-restoring counter terms can be computed using dimensional
reduction, \ie the $\drbar$ scheme. 

The fact that the $q$-$\sq$-$\tilde{\chi}$ couplings has a
supersymmetric limit means that we cannot numerically decouple the
gluino contribution from this process and obtain something like a
scalar leptoquark limit. While this limit is perfectly renormalizable,
a very heavy gluino breaks supersymmetry and the ultraviolet structure
of the coupling. We have checked that to decouple the gluinos we have
to absorb a logarithm $\log \mgo$ into an explicit
decoupling~\cite{decouple}, in analogy to
Eq.~\ref{eq:alphas_ct}.

%%%%%%%%%%%%%%%%%%%%%%%%%%%%%%%%%%%%%%%%%%%%%%%%%%%%%%%%%%%%%%%%%%%%%%%%
\subsection{On-shell subtraction}
\label{sec:on_shell}

%------------------------------------------------
\begin{figure}[t]
\includegraphics[width=0.5\textwidth]{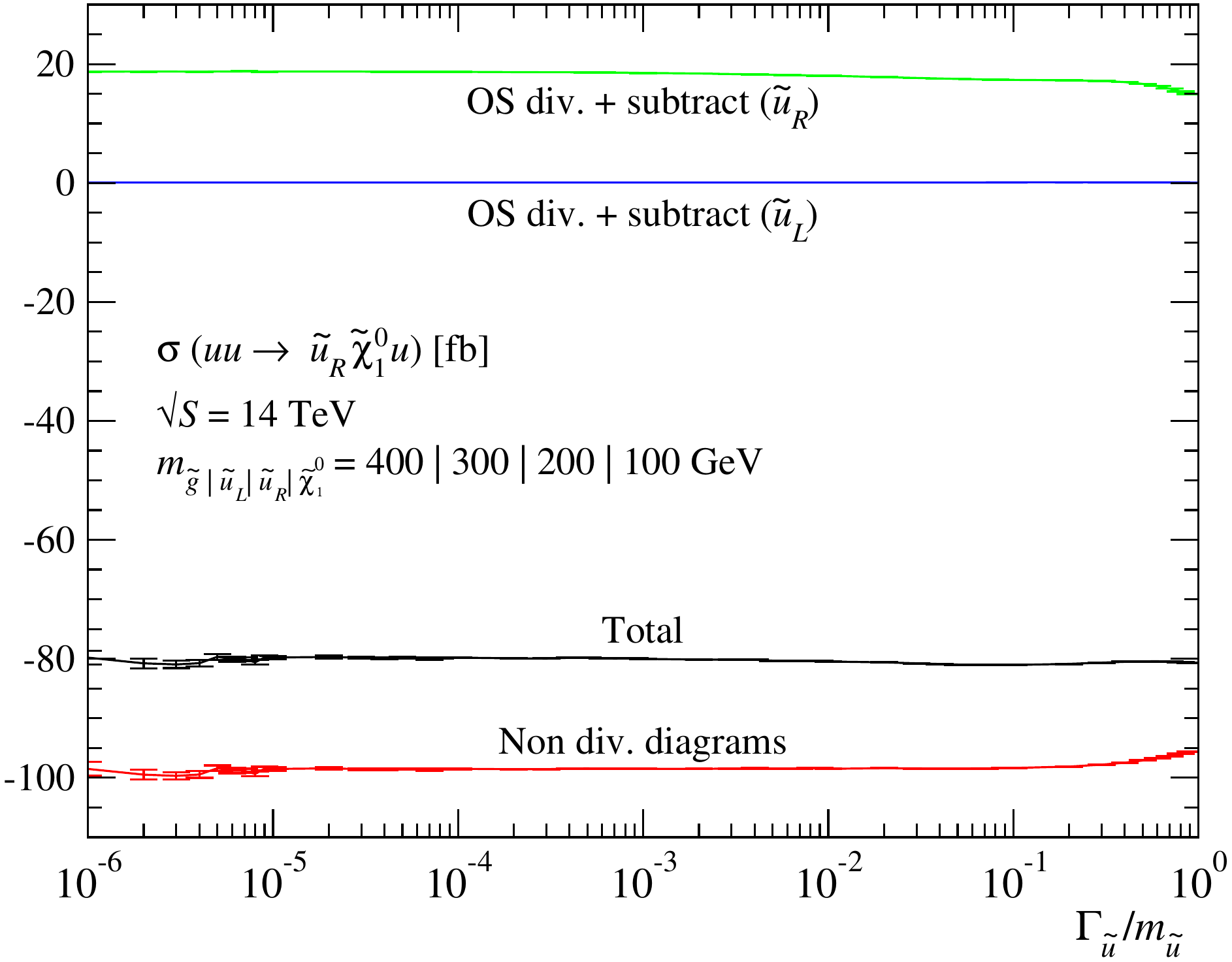} 
\caption{NLO contributions from intermediate on-shell particles
  contributing to $\sur \nz{1}$ production in the $u u$ initial
  state. We show the behavior as a function of the width-to-mass ratio $\Gamma_{\tilde{u}}/m_{\tilde{u}}$
of the on-shell squark, which acts as a mathematical cutoff in the {\sc Prospino}
subtraction scheme~\cite{prospino_onshell}, see the text for details. 
The masses are different from our usual benchmark point, to
illustrate all different channels.  The LHC energy is 14~TeV. Virtual corrections
  are not included.}
\label{fig:onshell}
\end{figure}
%------------------------------------------------

The Feynman diagrams in Fig.~\ref{fig:feyn2} naively lead to very
large corrections to the Born process. The reason is that for on-shell
intermediate states we would compare a $2 \to 2$ leading order process
proportional to $\alpha \alpha_s$ with a next-to-leading $2 \to 2$
contribution proportional to $\alpha_s^2$, multiplied by a branching
ratio which can be close to unity. However, the same diagrams are also
counted for example towards $\sq \sq^*$ production, including an
on-shell decay $\sq \to q \nz{1}$. To avoid double counting and to not
artificially ruin the convergence of the perturbative QCD description
of these production channels we remove all on-shell particle
contributions from the associated production, while the same Feynman
diagrams with off-shell propagators count towards $\sq \nz{1}$
production. To provide a reliable rate prediction, we need to define a
subtraction scheme which only removes the squared on-shell amplitudes
and which does this point by point over the entire phase space. This
{\sc Prospino} scheme~\cite{prospino_onshell} is defined as a
replacement of the Breit-Wigner propagator
\begin{equation}
 \frac{\matx(s_{q\tilde{\chi}})}{(s_{q\tilde{\chi}}-\msq^2)^2+\msq^2\Gamma_{\sq}^2} \longrightarrow  
 \frac{\matx(s_{q\tilde{\chi}})}{(s_{q\tilde{\chi}}-\msq^2)^2+\msq^2\Gamma_{\sq}^2} 
 \, - \, \frac{\matx(\msq^2)}{(s_{q\tilde{\chi}}-\msq^2)^2+\msq^2\Gamma_{\sq}^2}
 \; \Theta( \hat{s} - 4 \msq^2)
 \; \Theta( \msq - m_{\tilde{\chi}} ) \; .
\label{eq:sub1}
\end{equation}
In this form and for external squarks $\Gamma_{\sq}$ is only a
mathematical cutoff parameter. It can be chosen as their physical
width (as done in the {\sc Mc\@@Nlo} implementation~\cite{mcnlo}) or
as the uniquely defined and gauge invariant small-width limit
\begin{equation}
\lim_{\Gamma_{\sq} \ll \msq}
 \frac{1}{(s_{q\tilde{\chi}}-\msq^2)^2+\msq^2\Gamma_{\sq}^2} 
= \frac{\pi}{\msq \Gamma_{\sq}} \; \delta (s_{q\tilde{\chi}}-\msq^2) \; .
\label{eq:sub2}
\end{equation}
In {\sc MadGolem} this subtraction is automized. In
Fig.~\ref{fig:onshell} we show the numerical dependence and the
stability of our implementation of two simultaneous on-shell
subtractions, namely of $\sur \sur^*$ and $\sur \sul^*$,
production, where example Feynman diagrams are shown in
Fig.~\ref{fig:feyn2}. The intermediate gluino only appears in the 
$u\bar{u}$ process.
For values of $\Gamma/m < 10^{-2}$ 
the result is numerically stable, leading to
the limit shown in Eq.(\ref{eq:sub2}). It is obvious that including
the integrable interference terms with the continuum production is
necessary to obtain a stable and numerically correct result. This
correct description of the on-shell divergences also limits the
numerical impact of higher-order corrections to $\sul \nz{1}$
production proportional to the `wrong' $u$-$\sur$-$\nz{1}$
coupling.\bigskip

%------------------------------------------------
\begin{figure}[t]
\includegraphics[width=0.63\textwidth]{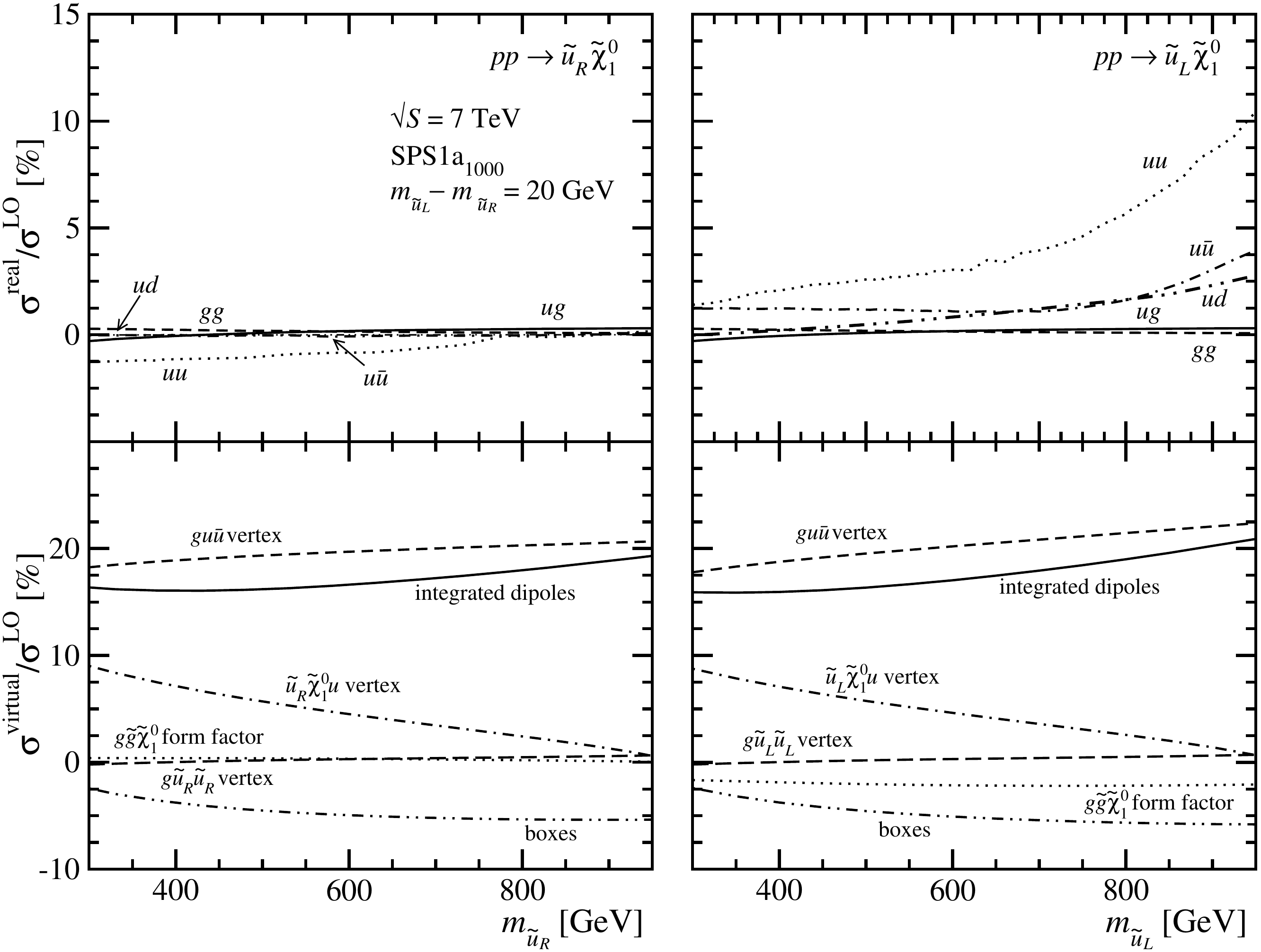} 
\caption{Relative size of the real and virtual corrections as a
  function of $m_{\tilde{u}_{R/L}}$ with a constant mass splitting
  $m_{\sul}-m_{\sur} = 20\,\gev$. Contributions from quark and squark self-energies
 lie below $1\%$ and are not explicitly shown.}
\label{fig:split}
\end{figure}
%------------------------------------------------

Once the on-shell diagrams are properly subtracted, we can analyze
the relative numerical impact of virtual and real corrections to $\sur
\, \nz{1}$ and $\sul \, \nz{1}$ production in
Fig.~\ref{fig:split}. The different contributions are defined through
dimensional regularization and (integrated) Catani-Seymour subtraction
terms with $\alpha=1$~\cite{catani_seymour}.

Genuine QCD effects constitute the dominant contribution. The bulk of
the virtual corrections arises from gluon-mediated $qqg$ vertex
corrections and the integrated dipoles. Each of them accounts for a
$+20\%$ shift in the cross section for our benchmark point. The
strongly suppressed gluino-mediated SUSY-QCD effects are explained by
the heavy mass suppression, effectively decoupling the gluino from the
theory. Correspondingly, the large vertex corrections are essentially
flat as a function of the squark mass.  The relative size of the integrated
dipoles slightly increase with growing $m_{\sq}$ towards the gluino
mass range.

Corrections to the $q$-$\sq$-$\nz{1}$ vertex can reach up to
$5\%$ for light squark masses.  The box diagram, in turn, gives a
negative contribution at the same level. Even milder is the effect
from the loop-induced $g$-$\go$-$\nz{1}$ form factor, but with a different
sign for the $\sul\nz{1}$ or the $\sur\nz{1}$ process.  This again
reflects the fact that they are sensitive to both the
$g_{u\sul\nz{1}}$ and $g_{u\sur\nz{1}}$ couplings and hence the
loop-induced $g$-$\go$-$\nz{1}$ form factor no longer factorizes with the
Born term.

%------------------------------------------------
\begin{figure}[b]
\includegraphics[width=0.7\textwidth]{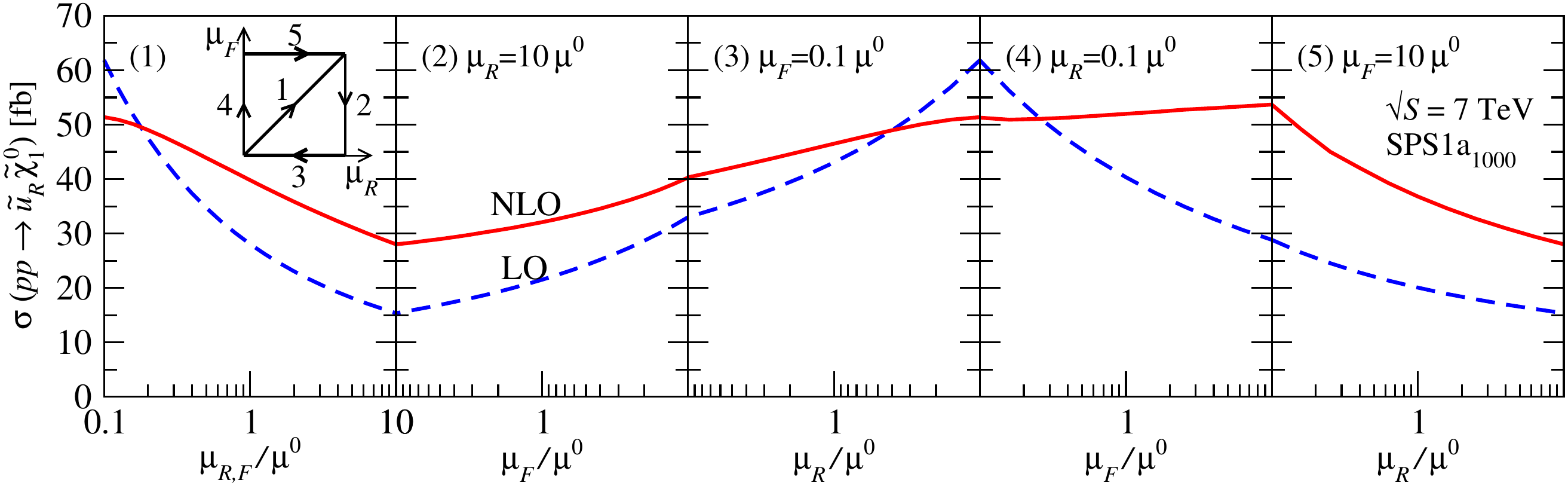} 
\caption{Profile of the renormalization and factorization scale
  dependence for $pp \to \sur \nz{1}$. The plot traces the scale
  dependence following a contour in the $\mu_R$-$\mu_F$ plane
  covering $\mu = (0.1 - 10) \mu^0$ as shown in the left panel.
  We assume our benchmark parameter choice and
  $\sqrt{S} = 7\,\tev$.}
\label{fig:scale1}
\end{figure}
%------------------------------------------------

%------------------------------------------------
\begin{figure}[t]
\includegraphics[width=0.5\textwidth]{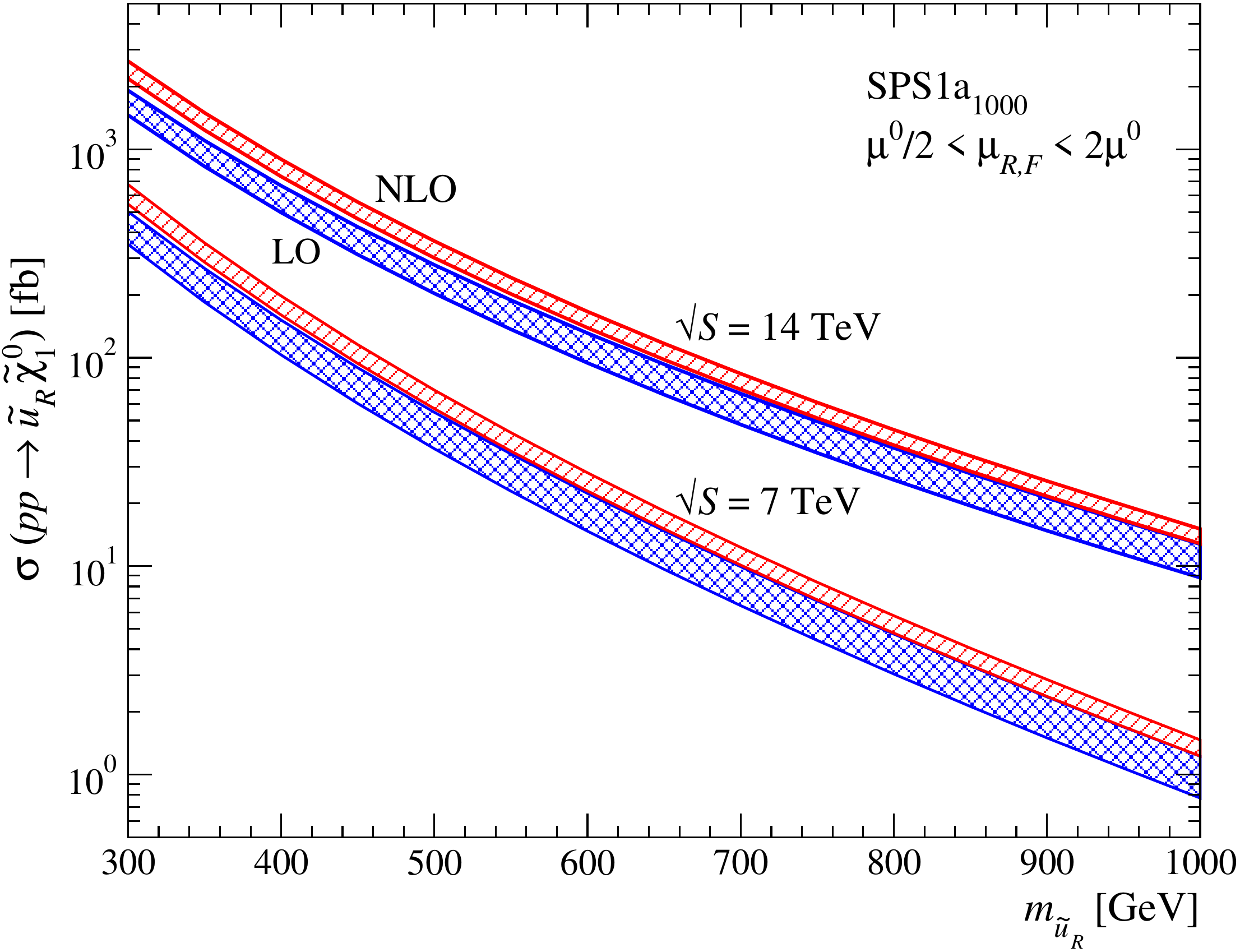} 
\caption{Total cross section for $pp \to \sur\,\nz{1}$ including the
  scale uncertainty. The band corresponds to a variation $\mu^0/2 <
  \mu_{R,F} < 2\mu^0$. We assume our
  benchmark parameter choice for all masses except for 
  $m_{\tilde{u}_R}$ and show results for $\sqrt{S} = 7$~TeV and 14~TeV.}
\label{fig:scale2}
\end{figure}
%------------------------------------------------

%%%%%%%%%%%%%%%%%%%%%%%%%%%%%%%%%%%%%%%%%%%%%%%%%%%%%%%%%%%%%%%%%%%%%%%%
\subsection{Scale dependences}

One of the main reasons to base LHC analyses on higher-order rate
predictions is the stabilization of the dependence on the unphysical
renormalization and the factorization scales. In Fig.~\ref{fig:scale1}
we present the cross sections as a function of the renormalization and
the factorization scales varied independently. We show the cross
section profile, both at leading order and at next-to-leading order,
moving along a contour in the $\mu_R$-$\mu_F$ plane. The contour is
defined in the left panel of Fig.~\ref{fig:scale1}. Individually
changing one of the two scales we expect a monotonous behavior, while
in the diagonal a distinct maximum appears for some processes. Again,
we only show the dominant $\sur\nz{1}$ channel for the modified SPS1a
benchmark point with a heavy gluino.
 
The stabilization becomes apparent in a considerable smoothing of the
$\sigma(\mu)$ slope in the entire $\mu_R$-$\mu_F$ plane.  This graphical
representation also indicates that the simultaneous variation of both
scales would have lead to a similar uncertainty estimate, but this
feature is clearly process and parameter dependent. Associated
production cross sections have the feature that they are
proportional to one power of $\alpha_s$ at leading order, which means
that unlike Drell-Yan-type channels the renormalization scale
dependence is visible at leading order. In contrast to QCD pair
production the renormalization scale does not completely dominate the combined scale
dependence, though. The NLO uncertainty band from the scale variation ranges
around $(\Delta \sigma)/\sigma \lesssim 20\%$, down from up to 70\% at
leading order.\bigskip

Finally, in Fig.\ref{fig:scale2} we show LO and NLO cross section
predictions including the scale dependence. We see that the two bands
are close to overlapping for $\mu^0/2 < \mu_{R,F} <
2\mu^0$, \ie our NLO result is within the LO error estimate; on
the other hand, we also see that a conservative scale variation should
not be chosen any smaller. Comparing the two LHC energies we see
that for 14~TeV the same number of signal events corresponds to an 
increase in the squark mass by at least 200~GeV.

%%%%%%%%%%%%%%%%%%%%%%%%%%%%%%%%%%%%%%%%%%%%%%%%%%%%%%%%%%%%%%%%%%%%%%%%
\subsection{MSSM parameter space}
\label{sec:mssm}

In Tab.~\ref{tab:sps} we survey the predictions for squark--LSP
production for all SPS points.  We define an inclusive $\sq\nz{1}$
cross section by including the first-generation squarks, $\sq =
\sul, \sur, \sdl, \sdr$, but without any approximation about the
individual squark masses. As we have shown, the contributions from the
second generation can be safely neglected. While the set of SPS
benchmark points is by no means a thorough coverage of the MSSM
parameter space, this brief scan is a useful starting point and will
help us understand effects from the MSSM parameter space. The whole
idea of the implementation of NLO corrections in the soon-to-be public
{\sc MadGolem} package is of course to allow for parameter scans by
any user.\bigskip

%------------------------------------------------
\begin{table}[t]
\begin{small}
\begin{tabular}{l||r||r|r|c||c|c|c|c||c|c|c|c} \hline
& $\sqrt{S} \, [\tev]$  & $\sigma^\text{LO}$ [fb] & $\sigma^\text{NLO}$ [fb] & $K$ & $K[{\sul}]$ & $K[{\sur}]$ & $K[{\sdl}]$ & $K[{\sdr}]$ & $m_{\su}$ & $m_{\sd}$ & $m_{\go}$ & $m_{\nz{1}}$ \\ \hline 
\multirow{2}{*}{\sps}& 7  & 35.27 & 50.44 & 1.43 & 1.52 & 1.42 & 1.46 & 1.47 & $\sul: 561$ & $\sdl: 568$  & \multirow{2}{*}{1000} & \multirow{2}{*}{97} \\
  & 14  &  215.02   & 301.27  & 1.40 & 1.49 & 1.39 & 1.42 & 1.43 & $\sur: 549$ & $\sdr: 545$ & & \\ \hline
\multirow{2}{*}{SPS1b}& 7  & 2.77 & 3.99 & 1.45 & 1.57 & 1.43 & 1.62 & 1.52 & $\sul: 872$ & $\sdl: 878$  & \multirow{2}{*}{938} & \multirow{2}{*}{162} \\
  & 14  &   27.21  &  37.46 & 1.38 & 1.48 & 1.36 & 1.52 & 1.43 & $\sur: 850$ & $\sdr: 843$ & & \\ \hline
\multirow{2}{*}{SPS2}& 7  & 0.04 & 0.07 & 1.52 & 1.81 & 1.49 & 1.69 & 1.65 & $\sul: 1554$ & $\sdl: 1559$  & \multirow{2}{*}{782} & \multirow{2}{*}{123} \\
  & 14  &   1.21  &  1.64 & 1.36 & 1.45 & 1.34 & 1.46 & 1.45 & $\sur: 1554$ & $\sdr: 1552$ & & \\ \hline
\multirow{2}{*}{SPS3}& 7  & 3.15 & 4.55 & 1.44 & 1.56 & 1.42 & 1.59 & 1.52 & $\sul: 854$ & $\sdl: 860$  & \multirow{2}{*}{935} & \multirow{2}{*}{161} \\
  & 14  &  30.20   & 41.59  & 1.38 & 1.49 & 1.36 & 1.50 & 1.43 & $\sur: 832$ & $\sdr: 824$ & & \\ \hline
\multirow{2}{*}{SPS4}& 7  & 6.44 & 9.04 & 1.40 & 1.52 & 1.38 & 1.53 & 1.49 &$\sul: 760$ & $\sdl: 766$  & \multirow{2}{*}{733} & \multirow{2}{*}{120} \\
  & 14  &   52.87  &  71.40 & 1.35 & 1.46 & 1.33 & 1.45 & 1.41 & $\sur: 748$ & $\sdr: 743$ & & \\ \hline
\multirow{2}{*}{SPS5}& 7  & 13.26 & 18.11 & 1.37 & 1.52 & 1.40 & 1.54 & 1.48 & $\sul: 675$ & $\sdl: 678$  & \multirow{2}{*}{722} & \multirow{2}{*}{120} \\
  & 14  &  95.81   & 132.29  & 1.38 & 1.50 & 1.37 & 1.49 & 1.43 & $\sur: 657$ & $\sdr: 652$ & & \\ \hline
\multirow{2}{*}{SPS6}& 7  & 9.84 & 14.06 & 1.43 & $\mathcal{O}(100)$ & 1.41 & 1.46 & 1.49 & $\sul: 670$ & $\sdl: 676$  & \multirow{2}{*}{720} & \multirow{2}{*}{190} \\
  & 14  &   77.08  &  107.03 & 1.39 & $\mathcal{O}(100)$ & 1.37 & 1.40 & 1.44 & $\sur: 660$ & $\sdr: 650$ & & \\ \hline
\multirow{2}{*}{SPS7}& 7  & 2.19 & 3.17 & 1.45 & 1.71 & 1.43 & 1.56 & 1.53 & $\sul: 896$ & $\sdl: 904$  & \multirow{2}{*}{950} & \multirow{2}{*}{163} \\
  & 14  & 22.36    & 30.80  & 1.38 & 1.61 & 1.36 & 1.46 & 1.43 & $\sur: 875$ & $\sdr: 870$ & & \\ \hline
\multirow{2}{*}{SPS8}& 7  & 0.65 & 0.95 & 1.45 & 1.66 & 1.43 & 1.62 & 1.57 & $\sul: 1113$ & $\sdl: 1122$  & \multirow{2}{*}{839} & \multirow{2}{*}{139} \\
  & 14  &  8.73  &  11.79 & 1.35 & 1.43 & 1.34 & 1.44 & 1.42 & $\sur: 1077$ & $\sdr: 1072$ & & \\ \hline
\multirow{2}{*}{SPS9}& 7  & 0.39 & 0.58 & 1.49 & 1.46 & $\mathcal{O}(1000)$ & 1.51 & $\mathcal{O}(1000)$ & $\sul: 1276$ & $\sdl: 1279$  & \multirow{2}{*}{1872} & \multirow{2}{*}{187} \\
  & 14  &  7.65   &  10.42 & 1.36 & 1.34 & $\mathcal{O}(1000)$ & 1.38 & $\mathcal{O}(1000)$ & $\sur: 1282$ & $\sdr: 1289$ & & \\ \hline
\end{tabular}
\end{small}
\caption{Summed cross section and corresponding $K$ factors for all four first-generation squark
  processes $pp \to \sq\nz{1}$ in different SPS benchmark
  scenarios. The scales are chosen at $\mu_{R,F}^0$.
  All masses are given in GeV.}
\label{tab:sps}
\end{table}
%------------------------------------------------
First, we see that the $K$ factors are largely insensitive to the
specific SPS point. This follows from the dominance of genuine QCD
effects, namely the gluon-mediated $u$-$u$-$g$ vertex
corrections and the real gluon emission. Changes in the squark and
gluino masses do not leave a recognizable fingerprint in the relative
size of the NLO corrections, which typically ranges around $40\%$.
The only exception are 
huge corrections for $\sul$ in SPS6 and 
$\sur,\sdr$ in SPS9, which are due
to essentially vanishing LO rates. To next-to-leading order rate
does not factorize with the LO couplings and is instead 
based on additional conjugate couplings.
Second, we see that the total cross sections do show a strong
correlation with the SPS points; the reason is twofold: on the one
hand there is kinematics, \ie the cross sections strongly depend on
the final state masses in phase space; in addition, we see dynamics
effects, where the strength of the $\nz{1} q \sq$ coupling changes
substantially from one scenario to another. For relative light spectra
in the range $m_{\sq} \sim 500 - 700 \,\gev$ and $m_{\nz{1}} \sim
100\,\gev$, as is the case for \sps, SPS5, and SPS6, the NLO cross
sections range around tens (hundreds) of femtobarns at $\sqrt{S} = 7
\, (14) \,\tev$.  For TeV-scale squark masses, as in SPS2, SPS8 or SPS9,
the cross sections stay below the femtobarn level.  The $K$ factors
remain around $+40\%$.  This simple pattern hardly depends on the
gluino mass, since the SUSY-QCD corrections are sub-leading.

%%%%%%%%%%%%%%%%%%%%%%%%%%%%%%%%%%%%%%%%%%%%%%%%%%%%%%%%%%%%%%%%%%%%%%%%
\section{Comparison with multi-jet merging}
\label{sec:distri}

While experimental analyses based on NLO cross section incorporate
significant improvements of the central values and the theory
uncertainties, we need to ensure that this picture also includes the
main distributions. From earlier studies we know that the transverse
momentum and rapidity distributions of the heavy particles are
relatively stable with respect to higher-order
corrections~\cite{prospino_squark,thesis}. Moreover, at least for the
production of heavy particles QCD jet radiation should be well
described by the parton shower, because the collinear approximation
includes sizeable $p_{T,j}$ relative to the masses in the final
state~\cite{qcd_radiation}. Nevertheless, we can check quantitatively
how well the NLO distributions from our fixed-order {\sc MadGolem}
computation agree with multi-jet merging~\cite{ckkw,mlm}. As a
comparison and to estimate the associated theory uncertainties we use
the {\sc Mlm} scheme with up to two hard jets, as implemented in {\sc
  Madgraph}. Any additional jets are well described by the parton
shower~\cite{qcd_radiation}.\bigskip

%------------------------------------------------
\begin{figure}[t]
\includegraphics[width=0.58\textwidth]{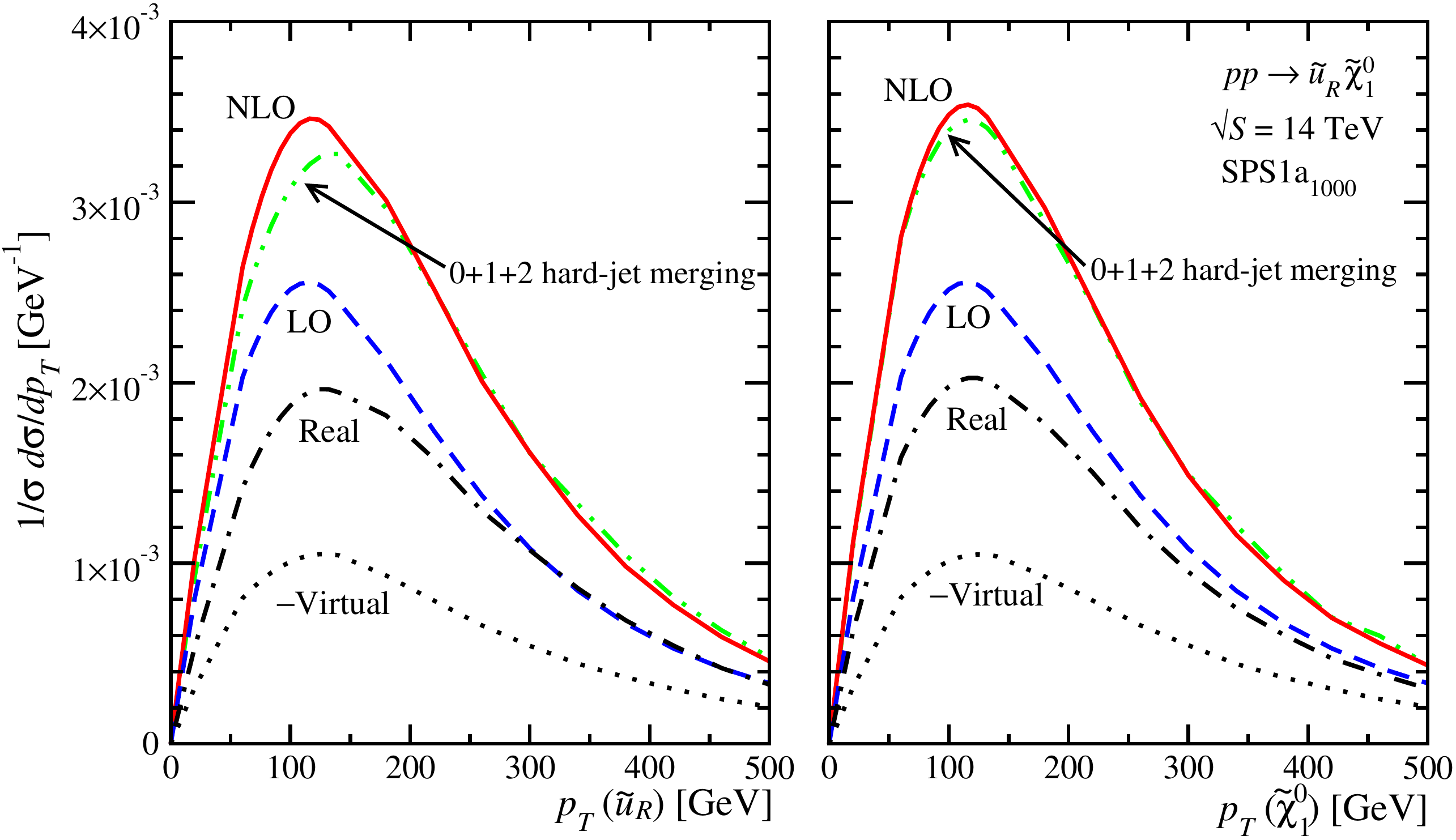}
\caption{Squark and neutralino $p_T$ distributions at the LHC
  ($\sqrt{S} = 14$~TeV) for \sps. We compare the merged
  sample with the fixed-order NLO computation. Both curves are
  normalized. We also show contributions to the NLO cross sections
  from the leading order, virtual and real parts. The latter are
  separated using the Catani-Seymour dipole with $\alpha=0.01$.}
\label{fig:distris}
\end{figure}
%------------------------------------------------

An important detail when using Catani-Seymour dipoles to compute
distributions is the assignment of the unintegrated and integrated
dipole contributions to phase space points. In the original paper,
Ref.~\cite{catani_seymour}, it is spelled out that the unintegrated
dipole contribution should not be counted towards their
$(n+1)$-particle or in our case 3-particle phase space point, but
towards the reduced 2-particle phase space. This ensures that for
example the $\alpha$ dependence of the integrated and unintegrated
dipoles exactly cancels not only for the total rate but also for all
distributions. On the other hand, this implies that the $p_{T,j}$
distribution remains unmodified from leading order and diverges at
small transverse momenta. To turn this distribution into a useful
prediction we need to consistently include a parton
shower~\cite{lecture}. Obviously, neither the leading jet
distributions nor the recoiling heavy system's distributions should be
used from the fixed order computation.

In Fig.~\ref{fig:distris} we show the transverse momenta of the squark
and the neutralino as computed by {\sc MadGolem} and using {\sc Mlm}
jet merging.  For the {\sc Mlm} simulation we cannot subtract the
on-shell singularities\footnote{In future releases, our {\sc Prospino}
  subtraction scheme should be available for {\sc Madgraph} at leading
  order.}. However, intermediate on-shell squarks with a subsequent
decay into a neutralino will produce considerably harder neutralinos
through the decay phase space. Therefore we only evaluate the dominant
$ug$ initial state. To the hard $ug \to \sur \nz{1}$ process we add
two hard and additional parton shower jets.

The normalized transverse momentum distributions agree well between
the next-to-leading order and the jet merging approaches; the slightly
harder result from jet merging can be attributed to additional recoil
jets. This is consistent with the observation that the real emission
contribution to the NLO result is slightly harder than its leading
order and virtual counterparts. In contrast to Fig.~\ref{fig:overmass}, where
we use $\alpha=1$, these distributions are computed with
$\alpha=0.01$, \ie introducing the unintegrated subtraction term only
very close to the soft and collinear poles. As a result, the virtual
corrections are now smaller than the real emission contributions. Any
physical observable is of course independent of the choice of
$\alpha$, as discussed above.

The normalization of the merged {\sc Mlm} sample is often an
improvement over the leading order result, but not in a consistent
quantifiable manner and to some degree dependent on the merging
parameters. Therefore, it should be adjusted to the consistent NLO
value.

%%%%%%%%%%%%%%%%%%%%%%%%%%%%%%%%%%%%%%%%%%%%%%%%%%%%%%%%%%%%%%%%%%%%%%%%
\section{Squark-neutralino channels}
\label{sec:channels}

%------------------------------------------------
\begin{figure}[t]
\includegraphics[width=0.5\textwidth]{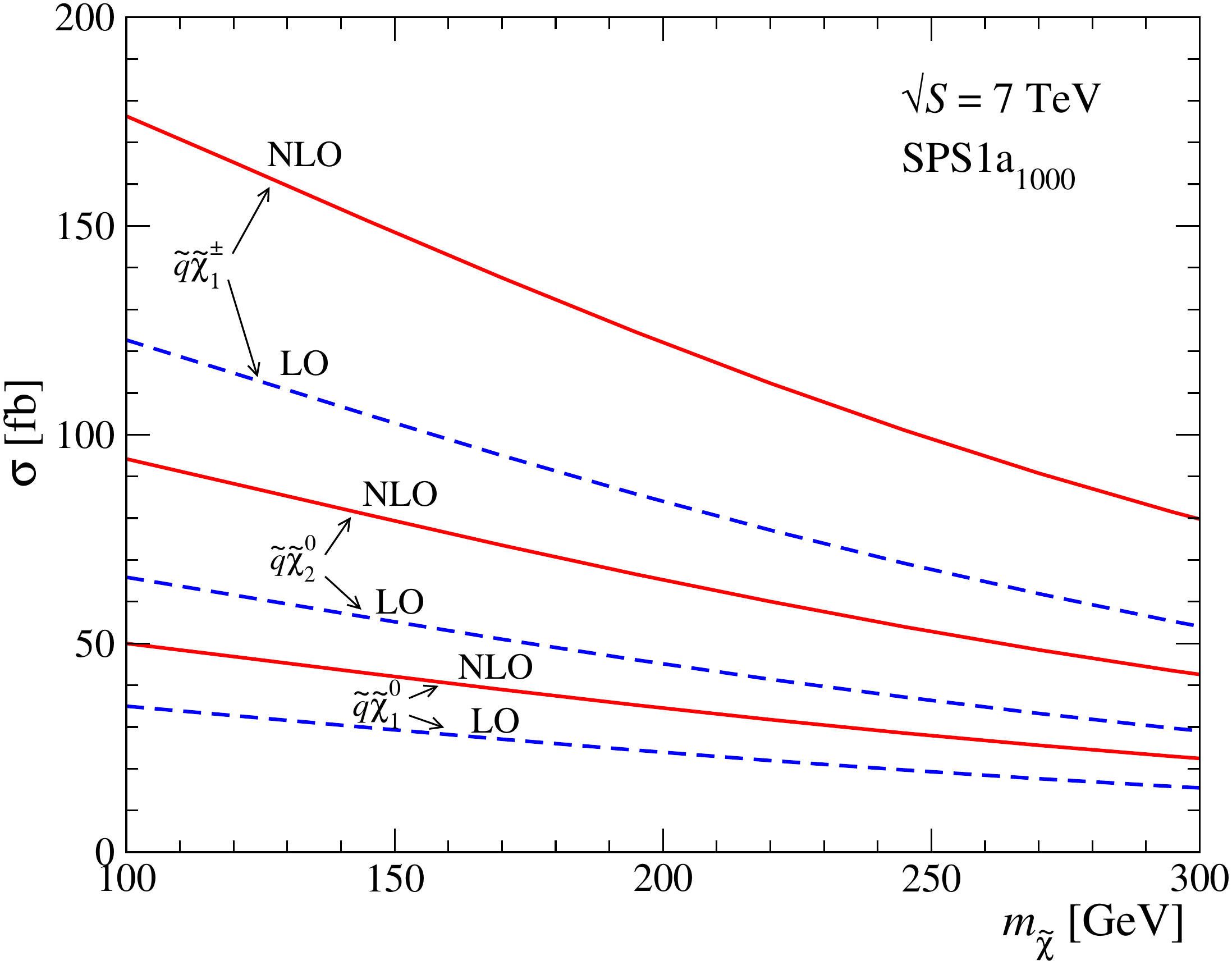}
\caption{Cross sections for different squark and neutralino/chargino
  production channels, $pp \to \sq\nz{1}, \sq \nz{2},
  \sq \cpm{1}$, as a function of the final-state
  neutralino/chargino mass.  We show results for $\sqrt{S} = 7\, \tev$
  and the modified SPS1a scenario.  As in Tab.~\ref{tab:sps} we sum
  over all first-generation squarks.  The scales are fixed to $\mu_{R,F}^0$.}
\label{fig:overm1}
\end{figure}
%------------------------------------------------

So far, we have concentrated in the associated production of a squark
with the lightest neutralino.  For example adding leptons to the
signature we would be sensitive to the production of a squark with
heavier neutralinos or charginos. On the level of the pure production
these channels often have larger rate than the bino-LSP channel.  In
Fig.~\ref{fig:overm1} we show the total cross sections for three
relatively large production channels: $pp \to \sq \nz{1}, \sq \nz{2},
\sq \cpm{1}$. The cross sections for both chargino charges are summed.
The differences in the total cross section for fixed
neutralino/chargino masses can be traced back to the size of the
$q$-$\sq$-$\tilde{\chi}$ couplings. While $g_{u\sur\nz{1}}$ is the
largest for the lightest neutralino, the coupling $g_{u\sul\nz{2}}$
dominates for the next-to-lightest neutralino.  The relative strength
$(g_{u\sul\nz{2}}/g_{u\sur\nz{1}})^2 \sim 1.8$ accounts for the
bulk of the difference between the predicted cross sections.  As
before, we emphasize that our computation is based on a completely
general MSSM input, encoded in a SLHA file~\cite{slha}. Negative mass
eigenvalues for the neutralinos can be
included~\cite{prospino_chargino}.

%%%%%%%%%%%%%%%%%%%%%%%%%%%%%%%%%%%%%%%%%%%%%%%%%%%%%%%%%%%%%%%%%%%%%%%%
\section{Summary}

We have computed the next-to-leading order QCD and SUSY-QCD
corrections to the associated production of a squark with a
gaugino. This channel could be responsible for the new physics
signature of one hard (decay) jet and missing transverse energy. This
computation is the first application of the fully automized and
soon-to-be public {\sc MadGolem} package. It makes no assumption about
the supersymmetric mass spectrum, for example of light-flavor squarks,
and relies on the SLHA interface of {\sc Madgraph}. Provided its QCD
counter terms are know, any new physics model with heavy strongly
interacting particles can be included into {\sc MadGolem} and its
relevant processes generated and computed.\bigskip

For the associated production of a squark and a neutralino we find
that NLO corrections due to gluon exchange and radiation dominate and
lead to a typical correction around $+40\%$. Gluinos only play an
appreciable role when they are very light, which in the light of
recent ATLAS and CMS measurements is becoming increasingly
unlikely. The NLO transverse momentum distributions we have compared
with a {\sc Mlm} merged computation and find good agreement for the
heavy particles produced. To avoid any dependence on merging
parameters, the total cross section should be based on the fixed-order
NLO computation.\bigskip

For associated production processes like the one considered here, it
is crucial to implement a consistent and stable on-shell subtraction
scheme separating associated production from QCD-mediated pair
production. The automization of this subtraction in the {\sc Prospino}
scheme is part of {\sc MadGolem}.

\acknowledgments First of all, we thank Steffen Schumann for his
advice on many QCD and simulation issues
and Rikkert Frederix for helping us to compare with {\sc MadFks}. 
Moreover, we are grateful to
the first author of Ref.~\cite{Allanach:2010pp} for outspokenly
insisting in a NLO rate prediction. 
DG acknowledges support by the International Max Planck Research School
for Precision Tests.
TP would like to also thank Wim
Beenakker, Michael Kr\"amer, Michael Spira, and Peter Zerwas for the collaboration
in an early stage of this project.
The work presented here has been in part supported by the Concerted
 Research action 
``Supersymmetric Models and their Signatures at the Large Hadron
 Collider'' of the Vrije Universiteit Brussel and 
by the Belgian Federal Science Policy Office through the Interuniversity
Attraction Pole IAP VI/11.

%%%%%%%%%%%%%%%%%%%%%%%%%%%%%%%%%%%%%%%%%%%%%%%%%%%%%%%%%%%%%%%%%%%%%%%%
\appendix
\section*{MadGolem: Automizing NLO predictions for new physics}

\begin{figure}[t]
 \begin{center}
  \includegraphics[width=0.8\textwidth]{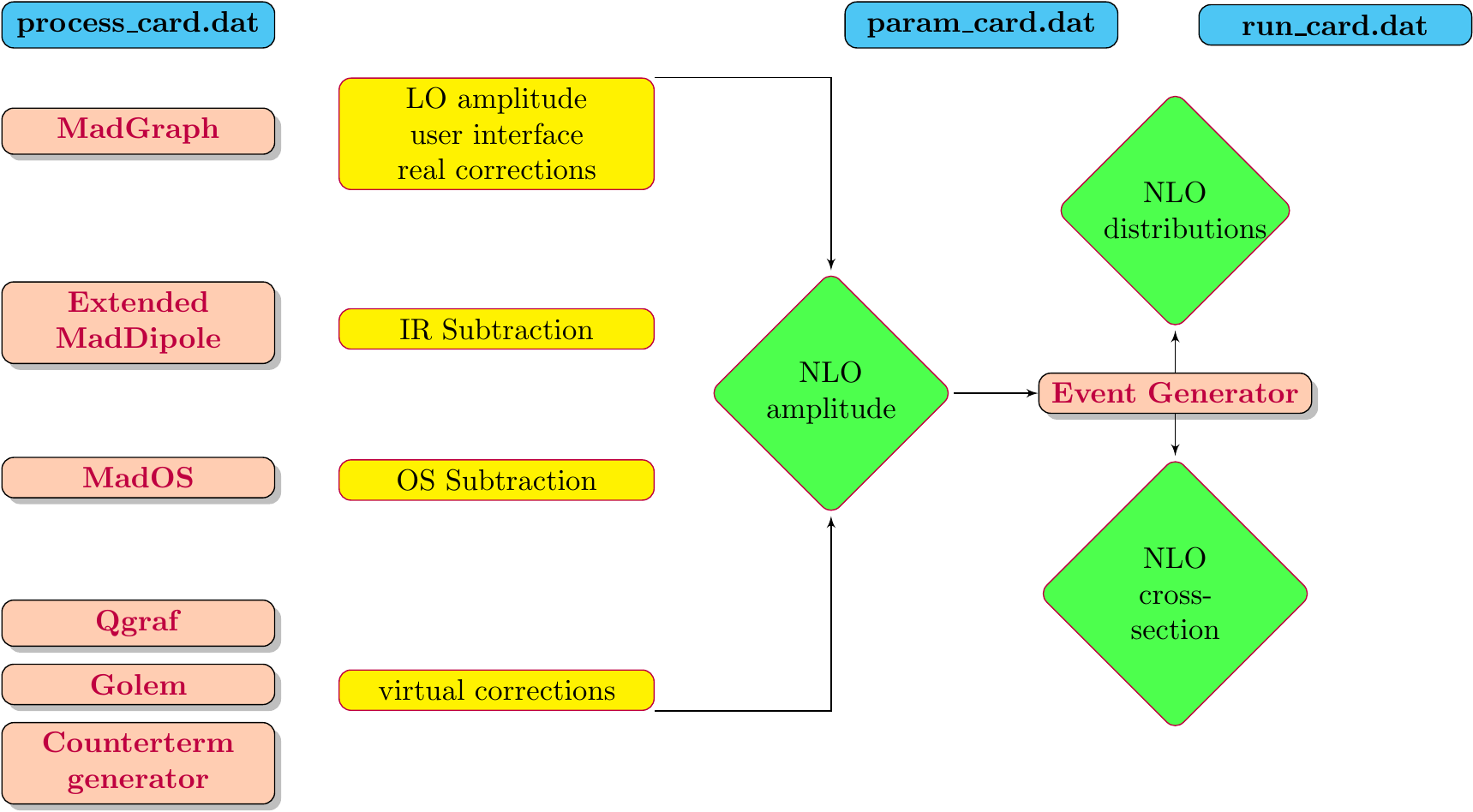}
\caption{Modular structure of {\sc MadGolem}.}
\label{fig:flowchart}
 \end{center}
\end{figure}

{\sc MadGolem} completely automates the calculation of cross sections
and the generation of parton-level events at NLO for arbitrary $2 \to
2$ processes in a generic new physics framework.  Its highly modular
structure we illustrate in Fig.~\ref{fig:flowchart}.  For the
generation of all tree-level Feynman diagrams and amplitudes we use
{\sc Madgraph}~\cite{madgraph}, which also provides user interfaces
and the basic code structure.  The one-loop Feynman diagrams we
generate with {\sc Qgraf}~\cite{qgraf} and a subsequent set of
specialized routines:
\begin{enumerate}
\item First, we translate the {\sc Qgraf} output into a code suitable
  for symbolic calculation languages.  The structures describing the
  Feynman diagrams and the corresponding Feynman rules we rewrite as
  algebraic expressions, keeping track of external wave functions,
  vertex couplings and internal propagators, color factors, Lorentz
  structure, and the overall sign from external fermion fields.  With
  this modification {\sc MadGolem} can cope with genuine features of
  new physics processes, such as Majorana fermions.
\item Then, we map the analytical evaluation of the color, helicity
  and tensor structures onto partial amplitudes, \ie a basis of color,
  helicity and tensor structures based on the spinor-helicity
  formalism.
\item In addition, we apply an analytical reduction to scalar loop
  integrals, based on a modified Passarino-Veltman reduction scheme
  implemented in {\sc Golem}~\cite{golem}.
\item Finally, we combine the results with the ultraviolet
  counter terms (which are also generated automatically). The final
  output for the virtual corrections we return both as analytical {\sc
    Maple} and numerical {\sc Fortran90} code.
\end{enumerate}

%------------------------------------------------
%------------------------------------------------

To compute complete NLO rates {\sc MadGolem} uses dedicated modules
for the different types of divergences appearing in NLO
calculations. To remove infrared divergences from real and virtual
gluon emission we resort to the Catani-Seymour dipole
subtraction~\cite{catani_seymour}. Our automatic dipole generation is
a modified and extended version of {\sc MadDipole}~\cite{maddipole},
now including all massive Standard Model and supersymmetric
dipoles. All integrated dipoles explicitly retain the dependence on
the phase space coverage of the subtraction term
$\alpha$~\cite{alpha}.  Intermediate on-shell divergences we remove
following the {\sc Prospino} scheme described in
Sec.~\ref{sec:on_shell}~\cite{prospino_onshell}. Its local subtraction
terms we generate automatically. The ultraviolet renormalization
counter terms we generate with {\sc Madgraph} amplitude. They are
expressed in terms of two-point functions which are supplied in a
separate library.  The current library supports QCD corrections for
arbitrary $2\ \to 2$ processes in the Standard Model and the MSSM and
can easily be generalized to other new physics models.\bigskip

{\sc MadGolem} does not require the user to interfere with the
computation of NLO cross section and parton-level events from the
basic input.  The model and the process are specified through {\sc
  Madgraph} input cards. Options like multi-particle notation are
supported together with additional specifications that allow us, for
instance, to separate QCD from SUSY-QCD contributions or retain
subsets of one-loop contributions.

We have performed exhaustive checks to ensure the reliability of {\sc
  MadGolem}.  Total NLO cross sections we have tested both in the SM
and the MSSM and covering numerous initial/final states, interactions,
and topologies. Cancellation of all divergences and gauge invariance
of the overall result have been confirmed numerically and
analytically.  The finite renormalized one-loop amplitudes we have
compared with {\sc FeynArts}, {\sc FormCalc} and {\sc LoopTools}
results~\cite{feynarts}. Particular attention we have paid to
numerical stability and convergence of our Catani-Seymour dipoles and
the on-shell subtraction.  Finally, we have checked our final results
with the literature (\eg $e^+e^- \to \sq^*\sq$~\cite{eesquark}) as
well as with {\sc Prospino}~\cite{prospino-slepton} and {\sc
  MadFks}~\cite{madfks} (\eg $pp \to
\tilde{\ell}^*\tilde{\ell}$).

%%%%%%%%%%%%%%%%%%%%%%%%%%%%%%%%%%%%%%%%%%%%%%%%%%%%%%%%%%%%%%%%%%%%%%%%

\end{document}